\documentclass[prl,reprint,superscriptaddress,preprintnumbers]{revtex4-1}

\usepackage{amsmath}
\usepackage{graphicx}
\usepackage{amssymb}
\usepackage{enumitem}
\usepackage{mathrsfs}
\usepackage[bbgreekl]{mathbbol}
\usepackage{amsmath,amssymb,amsfonts,amsthm}
\usepackage{mathrsfs}
\usepackage[linktocpage=true]{hyperref}
\usepackage{upgreek} 
\usepackage{titlesec}

\usepackage{color}

\newcommand{\st}{\sigma}
\newcommand{\ff}[1]{{\left\lfloor{#1}\right\rfloor}}

\newcommand{\kcs}{{\rm k}}

\newcommand{\nN}{{n_+}}
\newcommand{\nS}{{n_-}}

\def\nn{\nonumber}
\newcommand{\ii}{\mathrm{i}}
\newcommand{\ee}{\mathrm{e}}

\newcommand{\ugot}{u}

\newcommand{\f}[2]{{\frac{#1}{#2}}}

\newcommand{\p}[1]{{\left({#1}\right)}}
\newcommand{\comm}[1]{{\left[{#1}\right]}}

\newcommand{\spindle}{\mathbb{\Sigma}}

\newcommand{\dd}{{\rm d}}
\newcommand{\im}{{\rm i}}

\usepackage[dvipsnames]{xcolor}

\newcommand{\Zcm}{\zeta}
\newcommand{\nglobsym}{\mathfrak{d}}
\renewcommand{\upomega}{\epsilon} 
\newcommand{\numbchirals}{{|\mathcal{R}|}}
 \newcommand{\spm}{\sigma_\pm}

\newcommand{\pref}{\Psi}
\newcommand{\fks}{\sigma}
\newcommand{\bC}{\mathbb{C}}
\newcommand{\bZ}{\mathbb{Z}}
\newcommand{\fkn}{\mathfrak{n}}
\newcommand{\fkc}{\mathfrak{c}}
\newcommand{\fkb}{\mathfrak{b}}
\newcommand{\fkm}{\mathfrak{m}}
\newcommand{\fkr}{\mathfrak{r}}
\newcommand{\fkl}{\mathfrak{l}}
\newcommand{\cA}{\mathcal{A}}
\newcommand{\wti}{\widetilde\imath}
\newcommand{\wtj}{\widetilde\jmath}

\newcommand{\ys}{}

\begin{document}

\title{Microstates  of accelerating and supersymmetric\\[2.6mm]  AdS$_4$ black holes from the spindle index}

\titlespacing*{\section}{0pt}{15pt}{15pt}
\titlespacing*{\subsection}{0pt}{15pt}{15pt}

\author{Edoardo Colombo}
 \affiliation{Dipartimento di Matematica, 
 Universit\`a di Torino, 
Via Carlo Alberto 10, 10123 Torino, Italy}
\affiliation{INFN, Sezione di Torino,
 Via Pietro Giuria 1, 10125 Torino, Italy}
\author{Seyed Morteza Hosseini}
\affiliation{Department of Physics, Imperial College London, London, SW7 2AZ, UK}
\author{Dario Martelli}
 \affiliation{Dipartimento di Matematica, 
 Universit\`a di Torino, 
Via Carlo Alberto 10, 10123 Torino, Italy}
\affiliation{INFN, Sezione di Torino,
 Via Pietro Giuria 1, 10125 Torino, Italy}
\author{Antonio Pittelli}
 \affiliation{Dipartimento di Matematica, 
 Universit\`a di Torino, 
Via Carlo Alberto 10, 10123 Torino, Italy}
\affiliation{INFN, Sezione di Torino,
 Via Pietro Giuria 1, 10125 Torino, Italy}
\author{Alberto Zaffaroni}
\affiliation{Dipartimento di Fisica, Universit\`a di Milano-Bicocca,
Piazza della Scienza 3, 20126 Milano, Italy}
\affiliation{INFN, Sezione di Milano-Bicocca, Piazza della Scienza 3, 20126 Milano, Italy}

\begin{abstract}
\noindent  
We provide a first principles derivation of   the microscopic entropy of a very general  class of supersymmetric, rotating and accelerating black holes in AdS$_4$. 
This is achieved by  analysing the large-$N$ limit of the spindle index and completes 
the construction of the first example of a holographic duality involving supersymmetric field theories defined on orbifolds with conical singularities. 
\end{abstract}

\maketitle


\section{Introduction}\label{sec:intro}

 The explanation of the microscopic origin of the entropy of supersymmetric black holes in anti de Sitter (AdS) is one of the most spectacular 
 successes of the holographic duality. This was first accomplished in  \cite{Benini:2015eyy} for a class of AdS$_4$ black holes through the study of the large-$N$ limit of the topologically twisted index \cite{Benini:2015noa}.
The landscape of supersymmetric  black holes was significantly 
 broadened in  \cite{Ferrero:2020twa}, which 
constructed  a supersymmetric, rotating  and accelerating black hole 
with  \emph{spindle}
 horizon, displaying a number of remarkable features.
Most strikingly, in this solution supersymmetry is preserved via a novel mechanism, referred to as  anti-twist.
It is was later noted   
that supersymmetry on the spindle  may be preserved 
by means of a more  standard topological twist  \cite{Ferrero:2021etw,Ferrero:2021ovq}.
Utilising  the insight of \cite{Cabo-Bizet:2018ehj},
it was shown in   \cite{Cassani:2021dwa} that  the on-shell action of 
a supersymmetric and complex deformation of the black hole of \cite{Ferrero:2020twa} 
takes the form of an \emph{entropy function}, whose extremization yields  the Bekenstein-Hawking entropy.
A generalization of such entropy function was conjectured in \cite{Faedo:2021nub}, where it was proposed that it can be expressed in terms of
 gravitational blocks  \cite{Hosseini:2019iad}, as in all previous examples of black holes. The block decomposition of the gravitational entropy function
  was proved in    \cite{Boido:2022mbe}  using  the formalism of \cite{Couzens:2018wnk} and then in \cite{BenettiGenolini:2024kyy} employing equivariant localization in supergravity.

Motivated by these developments, \cite{Inglese:2023wky,Inglese:2023tyc} computed the 
localized partition function of ${\cal N} = 2$  Chern-Simons-matter theories defined on $\spindle \times S^1$, where $\spindle = \mathbb{WCP}^1_{[n_+,n_-]}$ is the spindle, 
with either twist or anti-twist for the $R$-symmetry connection $A$:
 \begin{align}
\label{twistantitwistdef}
 \int_{\spindle} \f{\dd A}{2\pi} = \frac{1}{2}\p{\frac{1}{\nS} +  \frac{\st}{\nN} } \equiv \frac{\chi_\st}{2} ~ , 
\end{align}
with $\st=\pm1$. The result can be expressed by a  single formula,  dubbed
  \emph{spindle index}  \cite{Inglese:2023wky},  which can be defined \cite{Cassani:2021dwa}  as a    flavoured Witten index   
\begin{align}
\label{traceformula}
Z_{\spindle \times S^1}
= {\rm Tr}_{\mathscr H\comm{\spindle}} \comm{\ee^{   - \im \sum_{\alpha=1}^{\nglobsym} \varphi_ \alpha Q_\alpha+ \ii \upomega J}}  \, , 
\end{align}
where $Q_\alpha$ are the generators of global symmetries of rank $\nglobsym$, $J$ generates  angular momentum on $\spindle$,  $\mathscr H\comm\spindle$ is the Hilbert space of BPS states on the spindle 
and the complex chemical potentials are related by the constraint
\begin{align}
\label{lovelyconstraint}
\sum_{\alpha=1}^{\nglobsym} \varphi_\alpha    + \frac{\chi_{-\st} }{2} \upomega  =   2 \pi n  \, , \qquad n \in \mathbb{Z} \,  .
\end{align}
In this letter we   will demonstrate that  the large-$N$ limit of the spindle index reproduces   the entropy functions associated to the supersymmetric and accelerating AdS$_4$ black holes. Explicitly, the entropy of a black hole with electric charges $Q_\alpha$ and angular momentum $J$ is obtained by extremizing  with respect to the variables $\varphi_\alpha$ and $\epsilon$ the entropy function
\begin{align} 
{\cal S}\, \equiv\, \log Z_{\spindle \times S^1} + \im \sum_{\alpha=1}^{\nglobsym} \varphi_ \alpha Q_\alpha- \ii \upomega J \, ,
\end{align}
under the constraint \eqref{lovelyconstraint}, setting $n=1$ and requiring that $J,Q_\alpha$ and ${\cal S}$ are real.
In the case $n_+=n_-=1$ our result encompasses  the large-$N$ limit of both the topologically twisted  index  and the generalized superconformal index. 
More details and  generalizations will be discussed in \cite{toappear}.

\section{The  spindle index matrix model}

We   consider  ${\cal N}=2$ Chern-Simons-matter quiver gauge theories with  
 gauge group $\mathcal G=\prod_{a=1}^{|\mathcal{G}|}$U$(N)_a$   and chiral multiplets transforming in either  bi-fundamental or adjoint representations of the gauge group factors. The index has been derived  using supersymmetric localization in \cite{Inglese:2023wky,Inglese:2023tyc} and it is written as the  matrix model
\begin{align}\label{eq: zs1spindlefull}
Z_{\spindle \times S^1}\p{\varphi,\mathfrak{n},\epsilon} =  \!\!\sum_{\mathfrak m \in \Gamma_{\mathfrak h} }\!\! \oint_{\mathcal C} \!\tfrac{\dd \ugot}{|W_{\mathcal G}|} \,  \widehat Z \p{\ugot, \mathfrak m |\varphi,\mathfrak{n},\epsilon }  ~ ,
\end{align}
where  $\mathfrak h$, $\Gamma_{\mathfrak h}$ and $W_{\mathcal G}$ denote   the Cartan algebra, the co-root lattice and the Weyl group of the gauge group $\mathcal G$, respectively; while  $\mathcal C$ is a suitable integration contour for $\ugot$. Here we have collectively expressed by  
$u\in\mathfrak h$ and $\mathfrak m\in\Gamma_{\mathfrak h}$
the gauge holonomies on $S^1$ and  fluxes through $\spindle$, respectively. 
Similarly,  $\varphi$ and $\mathfrak n$ are  flavour/topological charges and fluxes, with  (\ref{eq: zs1spindlefull})  implicitly depending    
on the spindle data $n_+,n_-$ and the twist parameter $\st$.

We focus on theories whose  gauge group and matter content can be represented by a quiver diagram with $|\mathcal G|$  nodes, 
where  an arrow   from node $a$ to node $b$  corresponds  to a bifundamental field in the representation $\mathbf{N}_{a} \otimes \overline{ \mathbf{N}}_{b}$, 
and  $a=b$ indicates  the adjoint representation.
For each U$(N)_a$ factor there are $N$ holonomies and fluxes, $(u_i^a,\mathfrak m_i^a)_{i=0}^{N-1}$; for each arrow we assign flavour charges and fluxes  $(\varphi_I,\mathfrak n_I)$,  where the index $I$ runs over \emph{all} the $\numbchirals$ chiral multiplets of the theory. If the corresponding arrow stretches from a node $a$ to a node $b$, 
we write $I\in(a,b)$. Moreover, for each node we assign charges/fluxes 
$(\varphi_m^a,\mathfrak n_m^a)$ for the topological symmetries.
As in \cite{Inglese:2023wky,Inglese:2023tyc} we consider a choice of R-symmetry that assigns even charges to the chiral multiplets: $r_I\in2\mathbb Z$.
For a chiral multiplet the corresponding chemical potential $\varphi_I$ is related to the flavour holonomy $u_I^F$ via 
\begin{align}
\varphi_{I}=2\pi u_{I}^F+ \left( \pi n-\frac\epsilon4 \chi_{-\st} \right) r_{I}\: ,
\end{align}
where, for each monomial term $W$ in the superpotential,
\begin{align}
  \sum_{I\in W} u_{I}^F= \sum_{I\in W} \mathfrak n_I =0 \, , \qquad  \sum_{I\in W} r_{I} =2 \: ,
\end{align}
so that 
\begin{align}
  \sum_{I\in W}\varphi_I + \frac{\chi_{-\sigma}}{2}\epsilon = 2\pi n  \: .
\end{align}
Notice that  the index $I$ runs over the fields belonging to a superpotential term, while in (\ref{lovelyconstraint}) the index $\alpha$ labels the generators of the global symmetries of the theory. For the ABJM model, that is the 
main focus in this letter, these two sets coincide. More general quivers will be discussed in \cite{toappear}.

The integrand of \eqref{eq: zs1spindlefull} is the product of a classical part and the 1-loop determinants of chiral and vector multiplets. In order to write it explicitly we need to introduce some further notation \cite{Inglese:2023wky}: first, we  define the symbols $\sigma_+=\sigma$ and $\sigma_-=-1$. Then, we set 
\begin{align}
\label{bc_def}
	\fkb_{ij}^{I}=&\,1-\frac{\fkm_i^{a}-\fkm_j^{b}}{n_+n_-}-\frac{\fkn_I}{n_+n_-}-\frac{r_I}2\chi_\fks
		-\cA^-_{I;\,ij}-\fks\:\cA^+_{I;\,ij}\:, \nonumber\\
	\fkc_{ij}^{I}=&\,\cA^-_{I;\,ij}-\fks\:\cA^+_{I;\,ij}\:, 
\end{align}
for each arrow $I\in(a,b)$, with
\begin{align}
\label{A_def}
	\fkl^\pm_{a;\,i} & =n_\pm\left\{\frac{\spm a_\pm\fkm^a_i}{n_\pm}\right\}\,	,\nonumber\\
	\cA^\pm_{I;\,ij}&= \,\left\{\frac{\fkl^\pm_{a;\,i}-\fkl^\pm_{b;\,j}+ \spm a_\pm \fkn_I - r_I/2}{n_\pm}\right\}\:,
	\end{align}
and   $a_\pm \in \mathbb{Z}$  such that  $n_+ a_- - n_- a_+=1$. Moreover $\{x\}\equiv x -  \ff{x}$. 
Notice that $\fkb_{ij}^{I}\in\mathbb Z$, while 
	 $\fkl_{a;\,i}^\pm,n_\pm \cA^\pm_{I;\,ij}\in \mathbb{Z}_{n_\pm}$.  
Denoting by 
\begin{align}
\label{y_def}
	& y_{ij}^{I}=\mathrm{e}^{-\ii \varphi_I-2\pi \ii (u_i^{a}-u_j^{b})}\cdot q^{\frac12\fkc_{ij}^{I}}~,   &  q=\mathrm{e}^{\im \epsilon}  ~ , 
\end{align}
the gauge holonomies, the 1-loop determinant contribution of the  chiral multiplets can be written as \cite{Inglese:2023wky,Inglese:2023tyc}
\begin{align}\label{CM1-loop}
	Z_{\text{1-L}}^{\rm CM} = \prod_{I=1}^{\numbchirals} \prod_{i,j=0}^{N-1} \Zcm^\sigma_q (y_{ij}^{I},\fkb_{ij}^{I} )~ ,
\end{align}
in terms of the function 
\begin{align}
\label{trick}
\Zcm^\sigma_q (y,\fkb) \equiv(-y)^{\frac{1-\fks-2\fkb}4}q^{\frac{(1-\fks)(\fkb-1)}{8}}
	\frac{(q^\frac{1+\fkb}{2} y^{-1};q)_\infty}
			{(q^{\frac{1-\fkb}{2\sigma}}   y^{-\fks};q)_\infty}\, , 
\end{align}
where 
$\p{z;q}_\infty$ is the    $q$-Pochhammer symbol, $y,q\in\bC$, $\fkb\in\bZ$ and $\fks=\pm1$. This  is 
the 1-loop determinant of a single chiral multiplet in an Abelian theory, satisfying 
\begin{align}\label{Zcm_symmetry}
	\Zcm^\fks_q(y,\fkb)= \Zcm_q^\fks(y^{-\fks},1-\fks-\fkb)^{-\fks}\, .
\end{align}
The 1-loop determinant  of all  vector multiplets reads
\begin{align}\label{VM1-loop}
	Z_{\text{1-L}}^{\rm VM } = \prod_{a=1}^{|\mathcal G|}  \prod_{i,j=0}^{N-1} \Zcm_q^\fks(y_{ij}^{a},\fkb_{ij}^{a})~ ,
\end{align}
where $y_{ij}^{a}$,  $\fkb^a_{ij}$, $\fkc^a_{ij}$,  and $\cA_{a;\,ij}^\pm$ are defined  as in \eqref{bc_def},
\eqref{A_def}, and \eqref{y_def}, with all the instances of $I$ and $b$ replaced by $a$, and with the following identifications:
$r_a\equiv2$, $\fkn_a\equiv0$, $\varphi_a\equiv2\pi n-\frac\epsilon2\chi_{-\st}$.

The classical part receives contributions from the Chern-Simons terms, which  can be written as \cite{toappear}
\begin{align}\label{eq: zcseff}
	Z^{\rm CS}_{\rm eff}  = \prod_{a=1}^{|{\mathcal G}|} \prod_{i=0}^{N-1}  \p{- y_{i}^{a}}^{\kcs_a \, (\fkb_{i}^{a} -1)} \, ,
\end{align}
where we  defined
\begin{align}
	y_i^{a}&=\mathrm{e}^{-2\pi \ii u_i^a}\cdot q^{\frac{\fkl^-_{a;\,i}}{2 n_-}-
		\sigma\frac{\fkl^+_{a;\,i}}{2 n_+}}\:, \nonumber\\
	\fkb_i^{a}&=1-\frac{\fkm_i^a}{n_+n_-}-\frac{\fkl^-_{a;\,i}}{n_-}-
		\sigma \frac{\fkl^+_{a;\,i}}{n_+}\:.
\end{align}
 In this paper we restrict to the case where  $\sum_a k_a=0$, corresponding to ${\cal N}=2$ Chern-Simons-matter quiver gauge theories with an M theory dual AdS$_4\times M_7$.
The topological symmetries also contribute to the classical part but the explicit expression will not be needed in this letter.

\section{Holomorphic block factorization}

The spindle index factorizes into the product of dual  holomorphic blocks \cite{Beem:2012mb}.
It is convenient to use a choice of factorization that breaks the Weyl-symmetry of the gauge group, generalizing the one introduced in
\cite{Choi:2019zpz} for the superconformal index. Starting from \eqref{CM1-loop}, we split the product over $i,j$ into a product over
$i<j$ and one over $i>j$, ignoring the diagonal terms that are subleading at large $N$; then we apply \eqref{Zcm_symmetry} to the $i>j$ terms and we find
\begin{align}
\label{factorization}
	Z_{\text{1-L}}^{\rm CM}\, = \, 
		\prod_{I=1}^{\numbchirals}\pref_I\cdot \mathcal B^+_I\cdot \mathcal B^-_I\:,
\end{align}
where for $I\in(a,b)$ we   defined
\begin{align}
\nonumber
	\pref_I&=\prod_{i<j}(y_{ij}^I)^{\frac{1-\fks-2\fkb^I_{ij}}4}(y_{ji}^I)^{-\frac{1-\fks-2\fkb^I_{ji}}4}\cdot q^{\frac{(1-\fks)(\fkb^I_{ij}-\fkb^I_{ji})}8}\:,\\
\label{B_def}
	\mathcal B^\pm_I&=\underset{\mathcal B^+:\:i>j}{\prod_{\mathcal B^-:\:i<j}}\frac
		{\p{\Big(\frac{z^\pm_{a;\,i}}{z^\pm_{b;\,j}}\Big)^{\pm\spm}\mathrm{e}^{-\ii\spm\Delta^\pm_I}q^{1-\cA^\pm_{I;\,ij}};q}_\infty}
		{\p{\Big(\frac{z^\pm_{a;\,j}}{z^\pm_{b;\,i}}\Big)^{\mp\spm}\mathrm{e}^{\ii\spm\Delta^\pm_I}q^{\cA^\pm_{I;\,ji}};q}_\infty}\:.
\end{align}
Notice that here we have swapped the role of $i,j$ in the blocks for convenience. The
$\pref_I$ will turn out to be subleading after the cancellation of the long-range forces.
The blocks $\mathcal B^\pm_I$ 
 depend on the combinations 
\begin{align}
\label{z_def}
\Delta^\pm_I & = \varphi_I \pm  \frac{\epsilon}{2} \left(\frac{{\mathfrak n}_I} { \nN \nS}  + \frac{\chi_{\st}}{2 }r_I \right) ~ ,\nn\\
 z^\pm_{a;\,i}  & =  \mathrm{e}^{\mp2\pi \ii u_i^a}\:q^{-\frac{\fkm_i^a}{2n_+n_-}} ~ .
	\end{align}
Notice that the variables $\Delta_I^\pm$ satisfy the constraints 
\begin{align}
	  \sum_{I\in W} \Delta_I^{\pm} = 2 \pi n + \frac{\spm\epsilon}{n_\pm} \, .
\end{align}
We derive the vector-multiplet counterparts of \eqref{factorization} and \eqref{B_def} by replacing the indices $I$ and $b$ with $a$ and applying standard identifications.

\section{Strategy for the large-$N$ limit}

We will implement the large-$N$ limit of the spindle index by relying on its factorization into holomorphic blocks,
generalizing the approach of \cite{Choi:2019zpz,Choi:2019dfu,Hosseini:2022vho}. 
For the partition functions on $S^2\times S^1$ the sum over all the possible values of the gauge fluxes 
$\fkm\in\Gamma_{\mathfrak h}\equiv\mathbb{Z}^{|\mathcal G|N}$
is usually approximated at large $N$ by promoting the fluxes $\fkm$ to continuous variables.
However, for $\spindle\times S^1$ this approximation is hindered
by the presence of  fractional parts in \eqref{A_def}.
To take care of this, we split each gauge flux as $\fkm_i^a\equiv n_+n_-(\fkm')_i^a+\fkr_i^a$, with $(\fkm')_i^a\in\mathbb Z$ and
$\fkr_i^a\in  \mathbb{Z}_{n_+n_-}$.  We then observe that there is a one-to-one correspondence between the possible values of $\fkl^\pm_{a;\,i}$ and
 $\fkr_i^a$:
\begin{align}\frac{ \fkl^\pm_{a;\,i}}{n_\pm}=\left\{\frac{\spm a_\pm\fkr^a_i}{n_\pm}\right\}\,  ,\,\,\,
	\frac{\fkr_i^a}{n_+n_-}=\left\{-\frac{\fkl^-_{a;\,i}}{n_-}-\fks\frac{\fkl^+_{a;\,i}}{n_+}\right\}\, .
\end{align}
We can therefore split the sum over $\fkm_i^a$ as
\begin{align}
	\sum_{\fkm_i^a\in\mathbb Z}=\sum_{\fkl_{a;\,i}^-=0}^{n_--1}\:\:\sum_{\fkl_{a;\,i}^+=0}^{n_+-1}\:\:\sum_{(\fkm')_i^a\in\mathbb Z}~ ,
\end{align}
and in the large-$N$ limit we  may  promote the $(\fkm')_i^a$ to be continuous variables while keeping the $\fkl^\pm_{a;\,i}$ discrete. Thus,  we approximate the integration measure of \eqref{eq: zs1spindlefull} by
\begin{align}
\label{large_N_measure}
	\sum_{\mathfrak m \in\Gamma_{\mathfrak h}}\!\! \oint_{\mathcal C} \!\tfrac{\dd \ugot}{|W_{\mathcal G}|} \,\, \longrightarrow \!\!\!\!
				\sum_{\fkl^\pm\in(\mathbb Z_{n_\pm})^{|\mathcal G|N}}
		\int_{\mathcal C^+}\dd z^+\int_{\mathcal C^-}\dd z^-,
\end{align}
at large $N$, where the variables $z^\pm$ were defined in \eqref{z_def}, $\mathcal C^\pm$ are appropriate
middle-dimensional contours in $\mathbb C^{|\mathcal G|N}$,
and the order of the Weyl group can be ignored since
$\log|W_{\mathcal G}|=\mathcal O(N\log N)$.

Since the $\mathcal B_I^\pm$ blocks depend separately on $z^\pm$, we will be able to perform the saddle point approximation in $z^-$ and $z^+$
independently of one another.
However the right hand side of \eqref{large_N_measure} also features a sum over the vectors of integers $\fkl^\pm$, which can take a total of 
$(n_+n_-)^{|\mathcal G|N}$ possible values, exponentially growing with $N$.
At large $N$ only one value of the $\fkl^\pm$ is expected to dominate at any given region of the parameter space: in particular, there will be one such value associated to the saddle point that reproduces the accelerating AdS$_4$ black holes. Two observations are in order to find the correct ansatz for $\fkl^\pm$:
first, we need to restrict our attention to the set of possible choices of $\fkl^\pm$ that, up to an appropriate permutation of the index $i$,
are periodic under shifts $i\to i+T$ for some $T\ll N$.
This assumption is necessary in order to be able to take (partially) the continuum limit: splitting the index $i$ as $i=Ti'+\wti$ with $\wti\in\{0,\ldots,T-1\}$, makes 
the fluxes $\fkl_{a;\,i}^\pm$ only depend on the index $\wti$, $\fkl_{a;\,i}^\pm\equiv \fkl_{a;\,\wti}^\pm$.
Hence,  at large $N$  the index $i'$ can be replaced with a continuous variable $t$.
Second, all the known methods for computing 3d partition functions at large $N$ \cite{Herzog:2010hf,Martelli:2011qj,Benini:2015eyy} 
require that terms with $i\sim j$ dominate
over the terms with $| i - j | \gg 1$.
The latter are called ``long-range forces" and with the appropriate assumptions they cancel out at leading order,
at least for a class of quiver theories that we shall discuss momentarily.
The cancellation of long-range forces constrains the possible choices of $\fkl^\pm$, although in general the constraint is complicated and it involves
the value of $z^\pm$ as well. Remarkably, the special value
\begin{align}
\label{fkl_ansatz}
	\fkl^\pm_{a;\,i}\,= \, i \mod \,\,  n_\pm 
\end{align}
makes the long-range forces vanish for any $z^\pm$.
We also anticipate that (\ref{fkl_ansatz}), along with a simple ansatz for $z^\pm$, reproduces the entropy of accelerating AdS$_4$ black holes.
Curiously, \eqref{fkl_ansatz} exhibits a strong  similarity to the ansatz  reproducing the entropy of AdS$_5$ black holes with arbitrary momenta, as discussed in \cite{Benini:2020gjh,Colombo:2021kbb}.

\section {Long-range forces cancellation}

In \eqref{B_def} the prefactors   $\pref_I$  encode  long-range forces  among the variables $z^\pm$ that could spoil the large-$N$ limit. As in previous work on $3d$ theories,
 we cancel the long-range forces by restricting to  ``non-chiral'' quivers, where for any bi-fundamental connecting the nodes $a$ and $b$ there is  a bi-fundamental connecting $b$ and $a$ and
\begin{align}\label{ABJ}
	\sum_{I\in(a)}  \mathfrak{n}_I = \sum_{ I\in(a)}  u^F_I =  \sum_{ I\in(a)}  (r_I-1) +2=0\,,
\end{align}
at each node $a$, where the sum is taken over all the arrows in the quiver with an endpoint at the node $a$, 
with adjoint chirals counting twice.
In a four-dimensional quiver this condition would be equivalent to the absence of ABJ anomalies  
for any symmetry. The conditions \eqref{ABJ} also imply that Tr $Q = 0$  for any global or $R$-symmetry symmetry with generator $Q$, where the trace is taken over all the fermions in the theory.

Using the periodicity relation $\fkl_{a;\,i}^\pm=\fkl_{a;\,i+T}^\pm$ that we have assumed, for the non-chiral quivers
satisfying \eqref{ABJ}
the product of all the prefactor terms \eqref{B_def} 
at large $N$ can be simplified down to
\begin{align}
\label{long_range_simplified}
	\prod_{I=1}^{\numbchirals}\pref_I\cdot\prod_{a=1}^{|\mathcal G|}\pref_a
		\, \, \longrightarrow\, \, \prod_{I=1}^{\numbchirals}\widetilde\pref_I\cdot\prod_{a=1}^{|\mathcal G|}\widetilde\pref_a~ ,
\end{align}
where for $I\in(a,b)$
\begin{align}
	\widetilde\pref_I=\prod_{s=\pm}\prod_{i,j=0}^{N-1}\p{\frac{z^s_{a;\,i}}{z^s_{b;\,j}}}^{\frac{\fks_s}4\p{1-\frac1{n_s}-2\cA^s_{I;\,ij}}
		\cdot\,\text{sign}(i-j)}
\end{align}
and a similar definition holds for $\widetilde\pref_a$.
Requiring the right hand side of \eqref{long_range_simplified} to vanish yields a mixed constraint on $\fkl^\pm$ and $z^\pm$.
Crucially, the ansatz \eqref{fkl_ansatz} is the only one that satisfies the property
\begin{align}
	\frac1{n_\pm}\sum_{j=j_0}^{j_0+n_\pm-1}\cA_{I;\,ij}^\pm=\frac12\p{1-\frac1{n_\pm}}
\end{align}
(and a similar relation with $i$, $j$ inverted) ensuring that the long-range forces coming from \eqref{long_range_simplified}  vanish for any  $z^\pm$.
Thanks to the Weyl-symmetry breaking factorization that we have used, the blocks $\mathcal B^\pm_I$ will not produce
any long-range term at leading order, as will now show.

\section{Holomorphic blocks at large $N$}

In order to compute the large-$N$ limit of the blocks $\mathcal B_I^\pm$, we will first consider the usual  ansatz for the saddle point distribution
of $z^\pm$ \cite{Herzog:2010hf,Martelli:2011qj,Benini:2015eyy},
\begin{align}
	\log z^\pm_{a;\,i}
	= -\spm N^\alpha t_i \mp \ii \ys y_a^\pm(t_i)\:,
\end{align}
where $t_i$, $y_a(t_i)$ are real and are assumed to be ordered so that $t_i\leq t_j$ for $i<j$.
The power of $N$ must be set to $\alpha=\frac12$, otherwise the 1-loop contributions and the Chern-Simons terms 
would grow with a different power law at large $N$ and it would not be possible to find non-trivial critical points.
When we take the continuum limit we split the index $i\equiv Ti'+\wti$:   assuming that the eigenvalues $t_i$ conform to a single continuous distribution at large $N$ allows  to  make   the replacements $t_i\equiv t_{i'}\equiv t$ and  define the eigenvalue density $\rho^\pm(t)$ such that 
\begin{align}
	& \frac{1}{N}\sum_{i=0}^{N-1}\bullet\longrightarrow\frac1 T\sum_{\wti=0}^{T-1}\int\dd t\rho^\pm(t)\bullet\:, & \int\dd t\rho^\pm(t) = 1 .
\end{align}
Expanding the $q$-Pochhammer symbols in terms of polylogarithms at all orders in $\epsilon$ 
and taking the large-$N$ limit  of each term as in \cite{Herzog:2010hf,Martelli:2011qj,Benini:2015eyy}  yields
\begin{align}
	\log \mathcal B_I^\pm=&N^{\frac32}\sum_{k=0}^2\epsilon^{k-1}\frac{B_k}{k!}\,\frac{1}{T^2}\sum_{\wti,\wtj=0}^{T-1}\int\dd t\rho^\pm(t)^2 \cdot\\
	\nonumber
		&\cdot g_{3-k}(-\spm\ys \delta y^\pm_{ab}(t)-\spm\Delta_I^\pm-\epsilon\cA^\pm_{I;\,\wti\wtj})+o(N^{\frac32})\:,
\end{align}
with $B_k =  B_k\p{1}=\{1,\frac12,\frac16,\ldots\}$ and
\begin{align}
\label{g_def}
	  g_{k}(x)=\frac{(2\pi)^k}{k!}
		B_k\p{\frac x{2\pi}+\nu}\:,
\end{align}
where $B_k\p{w}$ are the Bernoulli polynomials  and the integer $\nu$ in \eqref{g_def} must be chosen so that 
$\text{Im}\p{\frac 1\epsilon}<\text{Im}\p{\frac 1\epsilon (\frac x{2\pi}+\nu)}<0$.
We are using the  notation $\delta y^\pm_{ab}(t)\equiv y^\pm_a(t)-y^\pm_b(t)$. 

\section {The Large-$N$ limit of the spindle index}

We assume that the index is dominated by the configuration \eqref{fkl_ansatz}, which leads to a consistent large-$N$ limit. 
The large-$N$ limit of the classical Chern-Simons terms simplifies to 
\begin{align}
  \log Z^{\rm CS}_{\rm eff}  = \ys N^{3/2} \sum_a  \sum_{s=\pm} \frac{\fks_s}{\epsilon} k_a \int \dd t \, t \rho^s(t) y_a^s(t) \,.
 \end{align}
Consistently with fact that for saddles with gravity duals flux quantization implies $N/(n_+n_-)\in \mathbb{N}$  \cite{Ferrero:2020twa},
we can take $T=n_+ n_-$. Moreover, in order to compare with the  black hole solutions, we need to take $n=1$ \cite{Cassani:2021dwa}.
Finally, we also need to choose a determination:
we assume that $\text{Im}\p{\frac1\epsilon}<\text{Im}\p{\frac1{2\pi\epsilon}(y^\pm_{ab}(t)+\varphi_I)}<0$.
After some algebra, the explicit expression for $\fkl^{\pm}_{a;\,i}$ and the conditions \eqref{ABJ} yield  
\begin{align}\label{gravblock}
	\log Z_{\spindle \times S^1} =  -\sum_{s=\pm} \sigma_s \frac{F(\rho^s,\delta y_{ab}^s,\Delta^s_I)}{\epsilon} 
\end{align}
with
 \begin{align}\label{blockF}
&\frac{F(\rho^{\pm},\delta y_{ab}^{\pm},\Delta_I^{\pm})} {N^{3/2}} =  -\ys\sum_a k_a \int \dd t \, t \rho^\pm(t) y_a^\pm(t) \nonumber \\
&+\sum_{I\in (a,b)}\int \dd t \rho^\pm(t)^2 G_3^\pm( \ys \delta y_{ab}^\pm(t)+\Delta_I^\pm) \, ,
\end{align}
where $G_3^\pm(x)=\frac 16 x(x - \sum_{I\in W} \Delta^\pm_I/2)(x - \sum_{I\in W} \Delta^\pm_I)$.
The functions $G_3^\pm(x)$ are obtained by $g_3(x)$ in the range $\text{Re}\, x\in[0,2\pi]$ by replacing all occurrences of  $\pi$ with $\sum_{I\in W} \Delta^\pm_I/2$. The two terms in \eqref{gravblock} depend on different variables and they can be extremized independently.

For example, for the ABJM theory dual to AdS$_4\times S^7$, with $|\mathcal{G}|=2$, Chern-Simons level $k_1=1$ and $k_2=-1$ and four bi-fundamental fields transforming as $\mathbf{N}_{1} \otimes \overline{ \mathbf{N}}_{2}$ for $I=1,2$ and as  $\mathbf{N}_{2} \otimes \overline{ \mathbf{N}}_{1}$ for $I=3,4$, we find
\begin{align}
&
\eqref{blockF}
=  \ys\int \dd t  t \rho^\pm \delta y_{21}^\pm-\frac 12\int \dd t (\rho^\pm)^2\Big (\sum_I \Delta_I^\pm (\delta y_{21}^\pm)^2  \nonumber \\ & -\ys 2(\Delta_1^\pm\Delta_2^\pm- \Delta_3^\pm\Delta_4^\pm)\delta y_{21}^\pm-\sum_{I<J<K} \Delta_I^\pm \Delta_J^\pm \Delta_K^\pm \Big )\, .
\end{align}
This functional coincides with the large-$N$ limit of the effective twisted superpotential for the ABJM theory derived in \cite{Benini:2015eyy}, expressed in terms  of $\pm$ quantities, and its extremization is  straightforward \footnote{The functional ${\cal W}$ written in \cite{Benini:2015eyy}  has a critical point under the condition $\sum_I \Delta_I =2\pi$. We are using a homogeneous form of ${\cal W}$ where this condition has been used to eliminate all occurrences of $\pi$. The variables $\Delta_I$ in \cite{Benini:2015eyy} correspond to our $\Delta^\pm_I|_{\epsilon=0}$.}. The explicit expressions for $\rho^\pm$ and $\delta y_{21}^\pm$ can be found for example in \cite[(2.70)-(2.75)]{Benini:2015eyy}. 
The critical value is
 \begin{align}
 & F(\rho^{\pm},\delta y_{21}^{\pm},\Delta_I^{\pm}) \Big |_{\text{crit}}= \frac 23 N^{3/2} \sqrt{2 \Delta_1^\pm \Delta_2^\pm \Delta_3^\pm \Delta_4^\pm} \, .
\end{align}
Using \eqref{gravblock} we recover the gravitational block form \cite{Faedo:2021nub} \footnote{ To compare with the formulas in \cite{Faedo:2021nub} we set  $n=1$ and identify the variables as follows: 
$\pi \Delta_i^\pm |_{\rm there} =\Delta_I^\mp|_{\rm here}$, $-2\pi \epsilon |_{\rm there} =\epsilon |_{\rm here} \, , \pi \varphi_i |_{\rm there} =\varphi_I |_{\rm here}$, $r_i |_{\rm there}=r_I |_{\rm here}$, $\fkn_i |_{\rm there}= \fkn_I |_{\rm here}/(n_+n_-) + r_I \chi_\sigma/2$.}
  of the entropy function obtained in  \cite{Cassani:2021dwa}  and more generally conjectured in \cite{Ferrero:2021ovq}. The density of eigenvalues $\rho^\pm$ also agrees with the gravitational analysis performed in \cite{Boido:2023ojv}.

We can extend the result to  more general quivers: indeed, the term of order zero of $F$ in the $\epsilon$ expansion coincides  with the large-$N$ limit of the effective twisted superpotential of the ${\cal N}=2$ theory \cite{Hosseini:2016tor} 
 \begin{align}\label{twistedW}
&\ii \frac{{\cal W}(\rho,\delta y_{ab},\Delta_I)} {N^{3/2}} =  -\ys\sum_a k_a \int \dd t \, t \rho(t) y_a(t) \nonumber \\ 
&+\sum_{I\in (a,b)}\int \dd t \rho(t)^2 g_3(\ys \delta y_{ab}(t)+\Delta_I) \, ,
\end{align}
where $\Delta_I= \Delta^\pm_I|_{\epsilon=0}=2\pi u_I^F+\pi  r_I$
and we are ignoring topological symmetries for simplicity.  
This agrees  with well-known asymptotic behaviour of the holomorphic blocks \cite{Beem:2012mb}: $\log$(block)$= \ii\frac{{\cal W}}{\epsilon} +O(\epsilon)$. 
We then observe that \eqref{blockF} is a {\it homogeneous} form of the large-$N$ limit of the effective twisted superpotential ${\cal W}$ obtained by replacing $\Delta_I$  with $\Delta_I^\pm$ and all occurrences of $\pi$ with $\sum_{I\in W} \Delta_I^\pm/2$. The extremization of \eqref{blockF} is then equivalent to the extremization of ${\cal W}$ with 
the constraint $\sum_{I\in W} \Delta_I=2\pi$. One concludes that the entropy function has always a block form 
 \begin{align}\label{gravblock2}
	\log Z_{\spindle \times S^1} =  \frac{F_{\text{crit}}(\Delta^-_I)}{\epsilon} -\sigma \frac{F_{\text{crit}}(\Delta^+_I)}{\epsilon}\, .
\end{align}
We also see that the block function  $F_{\text{crit}}(\Delta_I)$, up to factors, is the homogeneous form of the large-$N$ on-shell value of the effective twisted superpotential ${\cal W}$. This has been computed  for many examples in \cite{Hosseini:2016ume}. At large $N$ ${\cal W}$  coincides with the $S^3$ partition function of the ${\cal N}=2$ theory \cite{Hosseini:2016tor} and, for theories with an AdS$_4\times M_7$ dual, the latter is in turn related  \cite{Herzog:2010hf,Martelli:2011qj} to the Sasakian volume \cite{Martelli:2005tp,Martelli:2006yb} of $M_7$.  Using this chain of equalities, one provides a field theory derivation of the gravitational block decomposition obtained in \cite{Boido:2022iye,Boido:2022mbe,BenettiGenolini:2024kyy} for configurations with a ``mesonic'' (or, ``flavour'') twist \cite{Hosseini:2019ddy}. 
More details about topological symmetries and issues with baryonic symmetries will be discussed in \cite{toappear}.

\section{Discussion}

 In this letter  we solved the fundamental 
 problem of elucidating  the microscopic origin of the Bekenstein-Hawking entropy of the most general class of  rotating BPS black holes currently known in four dimensions. Specifically, our findings demonstrate that the  microstates contributing to the entropy of accelerating black holes in four-dimensional anti de Sitter space-time are precisely mirrored by 
the physical degrees of freedom characterizing  
 three-dimensional gauge theories quantized on a spindle. To successfully  solve this   problem  we developed  a novel 
 approach tailored  to deal with  the degrees of freedom   of gauge theories on orbifolds. This technique holds vast  potential impact as it applies to  
 supersymmetric systems in any number of dimensions, including  \emph{e.g.} three-dimensional orbifold partition functions \cite{Inglese:2023tyc} and
 four-dimensional orbifold indices \cite{Pittelli:2024ugf}. Our results complete the construction of the first   
duality between a gravitational theory and a quantum field theory defined  on an orbifold, paving the way for a re-energized research program in holography.

\section*{Ackowledgments}
\noindent 
The work of EC and DM is  supported in part
by a grant Trapezio (2023) of the Fondazione Compagnia di San Paolo. AZ is partially supported by the MUR-PRIN grant No. 2022NY2MXY.
EC, DM, AP and AZ acknowledge partial support by the INFN.
SMH is supported in part by the STFC Consolidated Grants ST/T000791/1 and ST/X000575/1.
\bibliography{spindly.bib}

\end{document}